\documentclass{article}
\usepackage[a4paper,top=3cm,bottom=2cm,left=3.5cm,right=3.5cm,marginparwidth=1.75cm]{geometry}
\usepackage{mathpazo}
\pdfoutput=1
\usepackage{float}
\usepackage{graphicx}

\usepackage{ragged2e}
\usepackage{bm}
\usepackage{amsmath}
\usepackage{tabularx,ragged2e,array}
\usepackage[utf8]{inputenc}
\usepackage[T1]{fontenc}
\usepackage{placeins}
\usepackage{siunitx}
\usepackage{hyperref}
\usepackage{authblk}
\usepackage{lineno}
\usepackage[
  style       = nature,
  autocite    = superscript,
  backend     = biber,
  sortcites   = true,
  sorting = none,
  style       = numeric,
]{biblatex}
\title{Effects of thermal inversion \\induced air pollution on COVID-19\footnote{We thank Jeffrey Wooldridge for advice on our two-stage procedure. We thank Yujie Wang for advice with the air pollution data.}}

\author[1,3]{Klauber, H.\thanks{Corresponding author, klauber@mcc-berlin.net}}
\author[1,2]{Koch, N.}
\author[1,3]{Kraus, S.}
\affil[1]{{\footnotesize Mercator Research Center on Global Commons and Climate Change (MCC)}}
\affil[2]{{\footnotesize Potsdam Institute for Climate Impact Research (PIK)}} 
\affil[3]{{\footnotesize Technical University of Berlin (TU)}} 
\begin{document}
\maketitle
\begin{refsection}

\begin{abstract}
\noindent Air pollution is a threat to human health, in particular since it aggravates respiratory diseases. Early COVID-19 outbreaks in Wuhan, China and Lombardy, Italy coincided with high levels of air pollution drawing attention to a potential role of particulate matter and other pollutants in infections and more severe outcomes of the new lung disease. Both air pollution and COVID-19 outcomes are driven by human mobility and economic activity leading to spurious correlations in regression estimates. We use district-level panel data from Belgium, Brazil, Germany, Italy, the UK, and the US to estimate the impact of daily variation in air pollution levels on COVID-19 infections and deaths. Using random variation in air pollution generated by thermal inversions, we rule out that changes in mobility and economic activity are driving the results. We find that a 1\%-increase in air pollution levels over the three preceding weeks leads to a 1.5\% increase in weekly cases. A 1\%-increase in air pollution over four weeks leads to 5.1\% more COVID-19 deaths. These results indicate that short-term measures to reduce air pollution can help mitigate the health damages of the virus. 
\end{abstract}

\vspace{0.3cm}
\section{Introduction}
The detrimental effects of air pollutants on human health are well documented and some reductions in human exposure to them can be achieved fast and at low-cost\autocite{landrigan2018}. Since COVID-19 is a respiratory disease, policy-makers are considering short-term measures against air pollution as a means to mitigate infections and severe outcomes from the disease. A UK House of Commons working group, the \textit{All Party Parliamentary Group Air Pollution}, has for instance launched a strategy to reduce COVID-19 risks associated with air pollution\autocite{allpartyparliamentarygroupairpollution2020}. In France, local politicians have called on the national government to take "emergency measures", for instance to create low emissions zones, limit the use of liquid manure, and replace old wood stoves\autocite{lemonde2020}.\\

Early research has shown an association between COVID-19 outcomes and long-term\autocite{wu2020,cole2020b} and short-term \autocite{ogen2020,persico2020,zhu2020,isphording2020,travaglio2020} air pollution exposure for specific individual regions or countries. Air pollution is not randomly assigned to places, but results from different dimensions of human behaviour, some of which cannot be measured. Therefore, any correlation between air pollution and COVID-19 outcomes is likely to be biased. Air pollution levels have fallen as a result of COVID-19 outbreaks and lockdowns \autocite{he2020,cole2020a}. This pollution reduction can be due to direct behaviour changes as a reaction to the health threat, such as less car use (reverse causality) or due to a third, omitted variable that is driving both air pollution and COVID-19 outcomes, such as policy changes that occur independent from local infection clusters. We expect these reductions in air pollution to lead to substantial improvements in general human health \textcite{cicala2020}, but a robust cross-country estimate of the effects of air pollution reductions on COVID-19 outcomes is still lacking.\\

Here, we use thermal inversions that trap air pollution as a natural experiment to estimate the links between short-term increases in air pollution and COVID-19 infections and deaths. Satellite imagery and climate models let us measure thermal inversions and air pollution at high spatial resolution. We compile data from all countries that publish COVID-19 infections or deaths at the district-level to match the scale of our local weather-driven natural experiments. Our baseline sample comprises a geographically diverse set of countries -- Belgium, Brazil, Germany, Great Britain, Italy and the US. Therefore, we expect our analysis to have higher external validity than individual-country studies.\\

\section{Methods summary}
We use quasi-experimental methods from panel data econometrics to estimate the effect of air pollution on the number of infections and deaths from COVID-19. Rather than building a statistical model that fits the evolution of the pandemic well, we aim at recovering a single parameter. To this end, we use a Poisson model with case and death counts as the outcomes and air pollution predicted by the strength of thermal inversions as the treatment. Similar "reduced-form" approaches have for instance been applied to study the effect of weather on influenza \autocite{barreca2012} and the effect of anti-contagion policies \autocite{hsiang2020} and weather \autocite{carleton2020} on COVID-19.\\

Since it is difficult to find an empirical context, where individuals are plausibly randomly treated by air pollution, our parameter of interest can be best recovered in an ecological study. We use thermal inversions that generate random temporal and spatial variation in air pollution as a natural experiment. These episodes of lower temperatures at the ground level than in higher pressure levels trap air pollution. Our estimates are exclusively based on shifts in air pollution that are caused by these inversion events. This directly addresses concerns about reverse causality. For this instrumental variable approach to also restrict treatment variation to plausibly random variation, we need to argue that after removing variation with additional variables and fixed effects there is no other causal channel from thermal inversions to COVID-19 outcomes. This is called the "exclusion restriction"\autocite{angrist1996}. Inversions do not impact human health directly, but there could be a number of threats to our research design, which we address as follows.\\

District-level data on cases and deaths allows us to control for time-invariant features, such as topography, institutional strength or social capital at a high geographical resolution. We do this with a vector of dummy variables ("fixed effects") for each district in the sample leaving out one. This means we are only using changes within the same district over time to estimate our effect. Thereby we rule out that differences between districts in baseline conditions that are due to a history of inversions and human sorting behaviour as a reaction to pollution are driving the effect. We also absorb time-varying factors at the country-month and week levels.\\

If, after removing fixed effects, there remained causal channels between thermal inversions and COVID-19 outcomes that do not run through air pollution changes exclusively, the exclusion restriction would be violated. This could be the case if people are more prone to using their cars on days with inversions or if people spend more time with others indoors as a result of the temperature changes that create thermal inversions. We use thermal inversions at night, since day-time inversions can sometimes be seen \autocite{sager2019} and, thus, may create a systematic link between inversions and outcomes that violates the exclusion restriction. We also control for weather variables, such as precipitation, ground temperature, wind and UV radiation that vary daily at the district-level.\\

Our measure of thermal inversions is based on data from the ERA5 climate model and as our air pollution variable we use fine particulate matter smaller
than 2.5 $\mu m$ in diameter ($PM_{2.5}$) measured by Aerosol Optical Depth on MODIS satellite images. %We focus on $PM_{2.5}$ because of its well-documented high morbidity and mortality risk.
Since COVID-19 outcomes are affected with time-lags, we estimate the effect of thermal inversion induced air pollution changes during the current and two preceding weeks on weekly cases and current and three preceding weeks on weekly deaths. We also investigate the daily evolution of cases and deaths. This allows us to show that our results are robust to estimating our effect only based on variations within districts in the same month and removing all factors that vary in the same state on a weekly level and in the same country on a daily level. These could be changes in policies, economic activity, medical system performance, and testing regimes that introduce trends in our model that bias the comparisons between treated and control units at any given day and in any given intensity and that fixed effect regressions implicitly control (see Methods Section \ref{ssec::gb} for a discussion of the common trends assumption).

\section{Results}
\subsection*{The effect of thermal inversions on ambient air pollution}
First, we present evidence that the relationship between thermal inversions and ambient air pollution is strong. Table 1 shows the estimated effect of an additional degree in inversion strength on the logged mean $PM_{2.5}$ concentration. Because we assume that infections and deaths may be impacted by pollution with varying time lags, we estimate two different First Stage specifications. The left panel of Table 1 presents the effect of inversions on average $PM_{2.5}$ concentrations in a three-week time window, while the right panel uses a four-week time window.\\

The positive and statistically significant coefficients as well as F-statistics well above 10 in Table 1, show that inversions increase ambient particulate matter pollution robustly across specifications. For instance, the coefficient in column (3), indicates that a 1-degree increase in inversion strength increases the average three-week $PM_{2.5}$ concentration by about 0.58\%. On average, the strength of an inversion on a single day equals 2.27. Our effect estimate therefore implies that an additional weekday with an inversion leads to a 3.95\% increase in the weekly $PM_{2.5}$ level ($0.0058 \cdot 2.27 \cdot 3$). We derive an effect of 3.42\% from column (6) ($0.0032 \cdot 2.67 \cdot 4$). The average daily inversion strength differs slightly, given the different country samples underlying the two panels of the table. These results are in line with available evidence. Estimated effects of inversion incidents on $PM_{2.5}$ and $PM_{10}$ range from 2 to 4\% when converted to the week-level\autocite{arceo2016, hicks2016, jans2018}. \textcite{sager2019} shows that the effect of a 1-degree change in inversion strength increases the daily $PM_2.5$ concentration by about 10.47\%. This corresponds to a 0.50\%-increase in the three-week and a 0.37\%-increase in the four-week pollution concentration, which is close to the effects estimated in Table 1.\\

% (See section XY for a detailed derivation of the magnitude comparison).

%Hicks : 0.50- 1.07% increase in PM10 per additional inversion day per month 
%Arceo: 3.4% increase in PM10 per additional inversion day per week
% Sager: 10.47% increase in PM2.5 per additional degree in inversion per day (1.81 = average inversion strength)
% Jans : 24.5% increase in PM10, 32.6% increase in PM2.5 per inversion per day

% Umrechung Wochen-Level - Durchschnitts-Inversion
%Hicks: 0.5*28/7 - 1*28/7 = 2-4% PM10
%Arceo: 3.4%
% PM10
%Jans: 24.5/7 ,  32.6/7   = 3.5% PM10, 4.7% PM2.5
%Table1 3-week: 0.58*2.269841*21/7 = 3.95%
%(2.269841 = average daily inversion strength)
%Table1 4-week: 0.32*2.662995*28/7 = 3.41%
%(2.447867 = average daily inversion strength)
%\\

%Sager: 10.47% per day and inversion degree PM2.5
%Table1 3-week: 0.58*21 = 12.18%

\begin{table}[!ht]
    \centering
\scriptsize{
    \begin{tabularx}{\textwidth}{@{}p{4cm}
    S[table-format=2.4]S[table-format=2.4]S[table-format=2.4]|S[table-format=2.4]S[table-format=2.4]S[table-format=2.4]@{}}
   \hline
& \multicolumn{6}{c}{\textbf{logged} $\bm{PM_{2.5}}$}\\
& \multicolumn{3}{c}{three-week window}& \multicolumn{3}{c}{four-week window}\\
&{(1)} & {(2)} & {(3)} & {(4)} & {(5)} & {(6)}\\
\hline
\textbf{inversion strength} &0.0053{$^{***}$}&0.0074{$^{***}$}&0.0058{$^{***}$}&0.0026{$^{***}$}&0.0031{$^{***}$}&0.0032{$^{***}$}\\
&{(}0.0006{)}&{(}0.0006{)} &{(}0.0006{)} &{(}0.0006{)} &{(}0.0007{)}&{(}0.0007{)}\\[0.2cm]
\hline
F-statistic& {93.51} & {163.95} & {104.39}  & {17.08} & {23.08}& {25.61}\\
Observations& {101,046}& {101,046}& {101,046}& {65,569}&{65,569}& {65,569}\\
Countries& \multicolumn{3}{c}{\footnotesize{BEL, BRA, DEU, GBR, ITA, USA}}& \multicolumn{3}{c}{\footnotesize{BEL, BRA, DEU, USA}}\\
\hline
weather controls & {yes}& {yes}&  {yes}& {yes}& {yes}& {yes}\\
containment controls   & & {yes}& {yes}& & {yes}& {yes}\\
health system controls & & & {yes}& & & {yes}\\
\hline
\end{tabularx}}
\justify{
\scriptsize{
In each panel control variables are added sequentially from left to right. The first set of controls contains weather variables only, the second set adds controls for COVID-related containment and closure policies (e.g. school closings and stay at home requirements), and the third set adds COVID-related health system policies (e.g. testing policies and contact tracing). All regressions include district, week and state-month fixed effects. Standard-errors clustered at the district level are in parentheses.\\ Signif. Codes: ***: 0.01, **: 0.05, *: 0.1}}
    \caption{The effect of thermal inversions on ambient air pollution}
\end{table}

\subsection*{The effect of ambient air pollution on COVID-19 outcomes}

\noindent Next, we estimate how inversion-driven increases in air pollution affect COVID-19 infections and deaths. The left panel of Table 2 presents coefficients from regressions of the accumulated number of infections on the average predicted $PM_{2.5}$ concentration over the preceding three weeks. The right panel presents coefficients of the accumulated number of deceased COVID-19 patients on the average predicted $PM_{2.5}$ concentration over the preceding four weeks. The reported coefficients represent elasticities, that is percentage changes in the dependent variable linked to a 1\%-increase in the $PM_{2.5}$ mean. For instance, column (3) indicates that with every 1\%-increase in $PM_{2.5}$ in the preceding three weeks, the case numbers grow by 1.478\%. Given that we control for the accumulated case numbers of the preceding week, this 1.478\%-change takes place over a 7-day period. Table 2 also points to mortality effects. Column (6) shows, that a 1\%-higher pollution level in the past four weeks is linked to 5.120\% higher death counts. Again, this relative change refers to the death count level seven days ago.\\

\noindent The effect on case numbers and deaths becomes evident only when including policy controls. Regulations that are to restrict the spreading of the virus as well as changes in testing regimes both likely have a strong effect on the number of COVID-19 infections and cases eventually registered. Moreover, the different magnitudes of the coefficients in the left and right panel, point to a change in the rate of survival in registered COVID-19 patients. If deaths increased exclusively because the total number of patients increased, we would expect the relative effects to be approximately equal in magnitude. The difference could be explained if air pollution increases the severity of the disease or if it increases the share of vulnerable people in the registered  patients. To ensure that the differences across panels is not simply caused by the differing samples used, we re-estimate the left panel regressions with the sample including only Belgium, Brazil, Germany, and the USA. The estimated effects are very similar in magnitude.\\

\begin{table}[!ht]
    \centering
\scriptsize{
    \begin{tabularx}{\textwidth}{@{}p{4.5cm}
    S[table-format=2.3]S[table-format=2.3]S[table-format=2.3]|S[table-format=2.3]S[table-format=2.3]S[table-format=2.3]@{}}
   \hline
& \multicolumn{3}{c}{\textbf{COVID-19 cases}}& \multicolumn{3}{c}{\textbf{COVID-19 deaths}}\\
& \multicolumn{3}{c}{three-week window}& \multicolumn{3}{c}{four-week window}\\
&{(1)} & {(2)} & {(3)} & {(4)} & {(5)} & {(6)}\\
\hline
\textbf{predicted logged} $\bm{PM_{2.5}}$ &-0.023&1.018{$^{***}$}&1.478{$^{***}$}&4.015 & 4.577{$^{***}$} & 5.120{$^{***}$} \\
&{(}0.317{)}&{(}0.189{)}&{(}0.276{)}&{(}2.607{)} & {(}1.741{)} & {(}1.632{)}\\[0.2cm]
\hline
Observations& {72,021}& {72,021}& {72,021}& {20,658}& {20,658}& {20,658}\\
Countries& \multicolumn{3}{c}{\footnotesize{BEL, BRA, DEU, GBR, ITA, USA}}& \multicolumn{3}{c}{\footnotesize{BEL, BRA, DEU, USA}}\\
\hline
weather controls & {yes}& {yes}&  {yes}& {yes}& {yes}& {yes}\\
containment controls   & & {yes}& {yes}& & {yes}& {yes}\\
health system controls & & & {yes}& & & {yes}\\
\hline
\end{tabularx}}
\justify{
\scriptsize{
In each panel control variables are added sequentially from left to right. The first set of controls contains weather variables only, the second set adds controls for COVID-related containment and closure policies (e.g. school closings and stay at home requirements), and the third set additionally adds COVID-related health system policies (e.g. testing policies and contact tracing). In the left panel we control for the accumulated case number of the preceding week, in the right panel we control for accumulated death number of the preceding week. All regressions include district, week and state-month fixed effects. Standard-errors clustered at the district level are in parentheses. Signif. Codes: ***: 0.01, **: 0.05, *: 0.1}}
\caption{The effect of ambient air pollution on COVID-19 outcomes}
\end{table}

\vspace{-0.2cm}
\subsection*{The timing of pollution effects on COVID-19 outcomes}

Our prior results suggest that a higher pollution exposure is linked to increases in COVID-19 infections and deaths. However, from the findings at the weekly level, we cannot infer how quickly these effects occur after pollution increases and whether this timing is plausible given existing knowledge about the disease. $PM_{2.5}$ exposure is linked to cardiovascular and respiratory comorbidities\autocite{landrigan2018} that are in turn associated with worse COVID-19 outcomes. Studies also show, that air pollution can hamper the immune response to infections\autocite{ciencewicki2007air, becker1999exposure, kaan2003interaction, lambert2003effect, lambert2003ultrafine} and propagate their transmission\autocite{chen2010ambient, peng2020effects, ye2016haze}. While we cannot provide insight into potential mechanisms that apply with regards to COVID-19, we can derive that, if the development of symptoms or even the probability of transmission was affected, changes in the case numbers are unlikely to occur in the first days immediately after the pollution incident. This is because the average incubation time equals five days\autocite{linton2020,lauer2020} and because testing and reporting introduce an additional time lag \autocite{abbott2020}. Moreover, we would expect positive effects roughly in a time frame of two weeks after the pollution incident, as the vast majority of infected people develops symptoms within 12 days\autocite{lauer2020}. To test these theoretical considerations and to robustify our prior results, we now turn to an analysis at the daily level.\\

Table 3 shows how the total number of COVID-19 cases is affected by the presence of a thermal inversion on the past 21 days. For instance, the first coefficient in column (1) represents how a 1-degree increase in inversion strength on the current day changes registered cases compared to those registered yesterday. The last coefficient in column (3) represents the respective effect of the same change in inversion strength 20 days ago. In turning to the daily level, we adopt more restrictive fixed effects that control for district-month, state-week and country-day specific changes in the virus development. Because our policy controls vary at the country-day level they are redundant in this specification and we include weather covariates only. Overall, the table indicates statistically significant increases in the registered COVID-19 cases with a delay of 5 to 16 days, which is consistent with theoretical considerations. The effect estimated most precisely occurs at the 14th lag. Note that we do not intend to actually pin down effects to specific days. Rather, we view our daily analysis as indicative of a time range when pollution effects occur.\\

%\noindent To compare the effects estimated in Table 3 with those in Table 2, we conduct a back-of-the envelope calculation. First, we sum up all coefficients in Table 3 yielding a value of 0.006. This value represents the daily changes in COVID-19 cases linked to a 1-degree increase in inversion strength on every of the 21 days. The effect of a 1-degree increase in inversion strength on the daily $PM_{2.5}$ average is 3.15\% (see Table X in the Appendix). Dividing 0.006 by this number yields the daily percentage change in cases linked to a 1\% increase in air pollution over the past three weeks ($0.006/0.0315 = 0.1904762\%$). Finally, to convert the daily percentage change to the weekly percentage change, we take it to the power of 7 ($1.001904762^7 = 1.01341$). We obtain an elasticity of 1.341 which is close to the earlier estimated 1.478.\\

To compare the effects estimated in Table 3 with those in Table 2, we conduct a back-of-the envelope calculation. Dividing the sum of all 21 coefficients in Table 3 by the effect of a 1-degree increase in inversion strength on the daily $PM_{2.5}$ average ($0.006/0.0315 = 0.1904762\%$) and taking it to the power of 7 ($1.001904762^7 = 1.01341$), we obtain an elasticity of 1.341 which is close to the earlier estimated 1.478.\\

\begin{table}
\begin{center}\scriptsize{
\begin{tabularx}{0.75\textwidth}{@{}p{4.5cm}
    S[table-format=2.4]S[table-format=2.4]S[table-format=2.4]@{}}
   \hline
& \multicolumn{3}{c}{\textbf{COVID-19 cases}}\\
& {(1)} & {(2)} & {(3)}\\
& {$week_{t = -1}$} & {$week_{t = -2}$} & {$week_{t = -3}$}\\
\hline
inversion strength  $day_{ t = 1}$ &-0.0000&  0.0003  &0.0012{$^{***}$}\\
  &{(}0.0004{)}  &{(}0.0004{)}  &{(}0.0004{)}\\[0.1cm]
inversion strength   $day_{t = 2}$&0.0003&0.0008{$^{*}$}&0.0006{$^{*}$}\\
  &{(}0.0005{)}  &{(}0.0005{)}&{(}0.0003{)}\\[0.1cm]
inversion strength   $day_{t = 3}$&0.0000&0.0004&-0.0008{$^{**}$}\\
  &{(}0.0005{)}&{(}0.0007{)} &{(}0.0003{)}\\[0.1cm]
inversion strength   $day_{t = 4}$&0.0007&0.0010{$^{*}$}&-0.0006{$^{*}$}\\
&{(}0.0005{)}&{(}0.0007{)} &{(}0.0003{)}\\[0.1cm]
inversion strength   $day_{t = 5}$&0.0011{$^{*}$}&0.0002& 0.0001\\
  &{(}0.0006{)}   &{(}0.0005{)} &{(}0.0003{)}\\[0.1cm]
inversion strength   $day_{t = 6}$&0.0003&0.0002&-0.0000\\
 &{(}0.0006{)}&{(}0.0004{)}&{(}0.0003{)}\\[0.1cm]
inversion strength   $day_{t = 7}$&-0.0002&0.0000&0.0004\\
  &{(}0.0005{)} &{(}0.0005{)} &{(}0.0003{)}\\[0.1cm]
\hline
Countries& \multicolumn{3}{l}{\footnotesize{BEL, BRA, DEU, GBR, ITA, USA}}\\
\hline
\end{tabularx}}
\end{center}
\setlength{\leftskip}{1.8cm}
\setlength{\rightskip}{1.8cm}
\scriptsize{
The regression includes district-month, state-week and country-day fixed effects. We control for weather covariates as well as accumulated case numbers of the preceding day. Standard-errors clustered at the district level are in parentheses. The sample size is 236,074.\\  Signif. Codes: ***: 0.01, **: 0.05, *: 0.1}

\setlength{\leftskip}{0pt}
\setlength{\rightskip}{0pt}
\caption{Pollution effects on COVID-19 case numbers by days}
\end{table}

\noindent We also estimate the daily inversion effects on COVID-19 deaths, now considering a 28-day lag structure. There is vast evidence on the short-term effects of PM2.5 pollution on cardiovascular and respiratory diseases.\autocite{landrigan2018} Given that these diseases are comorbidities increasing the death risks of COVID-19\autocite{yang2020prevalence}, contemporaneous shifts in air pollution could have short-term effects on deaths among COVID-19 patients. Again, given delays in reporting, we would expect such effects to show only after a few days. If, as the results in Table 3 suggest, the number of symptomatic cases increases as a result of air pollution episodes, we would expect death numbers to rise with a longer delay. The median number of days between first symptoms and death is 18\autocite{zhou2020}. Therefore, if air pollution on the current day leads already infected people to develop symptoms and newly infected to develop symptoms after the incubation time of about five days, we would expect to see effects mainly around the 18th and the 23nd lag or somewhat later due to additional administrative delays.\\

\noindent Column (4) in Table 4 shows statistically significant coefficients for the 22nd to 26th lag. While the observed effects concentrate on the fourth week after pollution exposure, we do observe an earlier increase at the 16th lag as well. These findings support the hypothesis that air pollution increases the number of symptomatic cases. However, we do not observe statistically significant coefficients in the first weeks which would have been indicative of contemporaneous pollution effects on COVID-19 deaths. To compare the magnitude of the estimated effects with those in Table 2, we again derive the approximate elasticity. Dividing the sum of all 28 coefficients by the daily pollution effect ($0.0249/0.0315 = 0.7904762 $) and taking it to the power of 7 ($1.007904762^7 = 1.056663$), we obtain an elasticity of 5.667. This order of magnitude deviates only slightly from the earlier estimated 5.120.\\

\begin{table}
\begin{center}\scriptsize{
\begin{tabularx}{0.85\textwidth}{@{}p{4.5cm}
    S[table-format=2.4]S[table-format=2.4]S[table-format=2.4]S[table-format=2.4]@{}}
   \hline
& \multicolumn{4}{c}{\textbf{COVID-19 deaths}}\\
& {(1)} & {(2)} & {(3)} & {(4)}\\
& {$week_{t = -1}$} & {$week_{t = -2}$} & {$week_{t = -3}$} & {$week_{t = -4}$}\\
\hline
inversion strength $day_{t = 1}$ &0.0019&0.0007&0.0006&0.0027{$^{***}$}\\
  &{(}0.0019{)} &{(}0.0014{)}  &{(}0.0012{)} &{(}0.0010{)}\\[0.1cm]
inversion strength $day_{t = 2}$ &0.0004&0.0014&{0.0030$^{**}$}&0.0021{$^{**}$}\\
  &{(}0.0024{)} &{(}0.0015{)}&{(}0.0012{)}&{(}0.0009{)}\\[0.1cm]
inversion strength $day_{t = 3}$ &0.0005&-0.0016&0.0010&0.0016{$^{*}$}\\
  &{(}0.0025{)}&{(}0.0014{)}&{(}0.0011{)}&{(}0.0008{)}\\[0.1cm]
inversion strength $day_{t = 4}$ &-0.0020&0.0007&-0.0002&0.0015\\
  &{(}0.0023{)}&{(}0.0016{)}  &{(}0.0012{)} &{(}0.0009{)}\\[0.1cm]
inversion strength $day_{t = 5}$ &0.0003&0.0020&0.0003&0.0025{$^{***}$}\\
  &{(}0.0018{)}&{(}0.0016{)}&{(}0.0011{)}&{(}0.0009{)}\\[0.1cm]
inversion strength $day_{t = 6}$ &0.0005&0.0009&0.0016&0.0016\\
  &{(}0.0019{)}  &{(}0.0014{)} &{(}0.0010{)}   &{(}0.0010{{)}}\\[0.1cm]
inversion strength $day_{t = 7}$ &-0.0012&0.0001&0.0010&0.0010\\
  &{(}0.0018{)}  &{(}0.0015{)}  &{(}0.0011{)}&{(}0.0008{)}\\[0.1cm]
  \hline
  Countries& \multicolumn{4}{l}{\footnotesize{BEL, BRA, DEU, USA}}\\
\hline
\end{tabularx}}
\end{center}
\setlength{\leftskip}{1.2cm}
\setlength{\rightskip}{1.5cm}
\scriptsize{
The regression includes district-month, state-week and country-day fixed effects. We control for weather covariates as well as accumulated death numbers of the preceding day. Standard-errors clustered at the district level are in parentheses. The sample size is 82,423.  Signif. Codes: ***: 0.01, **: 0.05, *: 0.1}

\setlength{\leftskip}{0pt}
\setlength{\rightskip}{0pt}
\caption{Pollution effects on COVID-19 deaths by days}
\end{table}

%%%%%%%%
\section{Discussion}
We find that air pollution episodes created by thermal inversion increase COVID-19 infections and deaths. A 1\%-increase in $PM_{2.5}$ levels over the three preceding weeks leads to 1.5\% more weekly cases. A 1\%-increase in $PM_{2.5}$ levels over the four preceding weeks leads to a 5.1\% increase in COVID-19 deaths. Thermal inversions also drive daily increases in cases and deaths that are consistent with common priors on time lags of the disease and its measurements.\\

Our results indicate that even short-term reductions in air pollution can help mitigate the spread and severity of COVID-19. Since reductions in air pollution are known to generate large net benefits, particularly in countries with high pollution levels, many short-term measures to curb air pollution are low regrets options, as long as they do not divert attention from the core measures needed to mitigate COVID-19 directly.\\

We only estimate the net short-term effect of air pollution on COVID-19 outcomes controlling for baseline differences between populations in mid- and long-run exposure. Therefore, it can be considered a lower bound for the potential benefits from reducing air pollution over longer time periods. Estimates of the heterogeneity of COVID-19 outcomes in terms of mid- and long-term exposure to air pollution are complementary to our approach and could help target policies at vulnerable populations, particularly if analyzed in interaction with short-term air pollution shocks.\\

Our reduced-form approach aims at getting as close as possible to a causal estimate of the effect of air pollution on COVID-19 outcomes. Our empirical strategy builds on previous studies in epidemiology\autocite{firket1931,firket1936,schrenk1949,seaton1995,anderson1995,anderson2009} and economics\autocite{hanna2015,arceo2016,hicks2016,jans2018,sager2019,salvo2017} that have identified thermal inversions as natural experiments for episodes of increased air pollution. However, it does not speak to the biological mechanisms that create the effect. To learn more about the underlying mechanisms, combining quasi-experimental variation in air pollution with individual level health observations might be helpful. Differences in air filtration in the work place or in schools created by regulations or renovation cycles could be a suitable setting for an analysis. 
\newpage
\printbibliography
% \section{Figures}
\end{refsection}

%%TC:ignore
\newpage
\begin{refsection}
\section{Methods}\label{sec::methods}
\subsection{COVID-19 data}
\noindent We compile daily data on COVID-19 cases and deaths at the district level. At the time of data collection in May/June 2020, we were able to find information at this fine spatial and temporal resolution for six countries: Belgium, Brazil, Germany, Italy, United Kingdom and United States. All data are openly available online and provided by official sources. The following table provides an overview of the data by country. All data were scraped on 9th June.\\

\noindent We collect the number of accumulated COVID-19 cases and deaths. While we observe case numbers for all six countries, death counts are available only for four of them.\footnote{Data for Belgium contain death counts only for Brussels.} The reporting of these numbers differs across and within countries. For instance, in some districts they include only lab-confirmed cases, while in others they also include probable cases. Moreover, the time of reporting may differ as well as the testing regimes within countries and districts over time. While we are not able to observe these information in our data, our empirical model addresses these differences with the inclusion of fixed effects.\\
\FloatBarrier

\begin{table}[!ht]
\begin{scriptsize}
\begin{tabular}{p{2.7cm}|p{1cm}p{1cm}p{5cm}p{3cm}}
\textbf{Country} & \textbf{Cases}  & \textbf{Deaths} & \textbf{Original Source} & \textbf{Reference} \\
\hline
 Belgium & yes & partly & Sciensano  &  \textcite{databe}  \\
Brazil & yes & yes &  Ministério da Saúde & \textcite{databr}   \\
Germany& yes & yes & Robert Koch-Institut&   \textcite{datade}\\ 
Italy & yes & no & Dept. of Civil Protection & \textcite{datait}\\
United Kingdom & yes & no & Dept. of Health and Social Care & \textcite{datauk}\\
United States & yes & yes & local governments and health departments&\textcite{dataus} \\
\end{tabular}\end{scriptsize}
\end{table}

\FloatBarrier

\subsection{Thermal inversions and weather data}\label{ssec::era5}
\noindent We obtain data on atmospheric air temperature as well as weather conditions from ECMWF (The European Centre for Medium-Range Weather Forecasts). The product ERA5 provides interpolated hourly data at 37 different pressure levels with a spatial resolution of 31km (0.28125 degrees).\\

\noindent We define inversions as episodes when the difference in temperature between the 925 hPa pressure level (ca. 600 m above sea level) and the 1000 hPa pressure level (ca. 30 m above sea-level) is positive. We weight inversions by their strength, that is the continuous difference between the two temperature layers. Following recent studies, we consider night-time inversions using temperature measurings at 2 a.m. local time \autocite{sager2019, jans2018}.\\

\noindent We aggregate the gridded data to the district level (GADM-GID2) by calculating the weighted mean, where the weights equal the fraction of the grid covered by the district. For our analysis at the weekly level, we sum up all inversions within moving three-week and four week time windows. Figure 1 shows that inversion episodes occur with great temporal and spatial variation across districts.\\

\begin{figure}[!ht]
        \begin{center}
             \caption{Spatial and temporal variation in the occurrence of inversion episodes}
        \includegraphics[clip = 5 0 5 0, width = 0.8\textwidth]{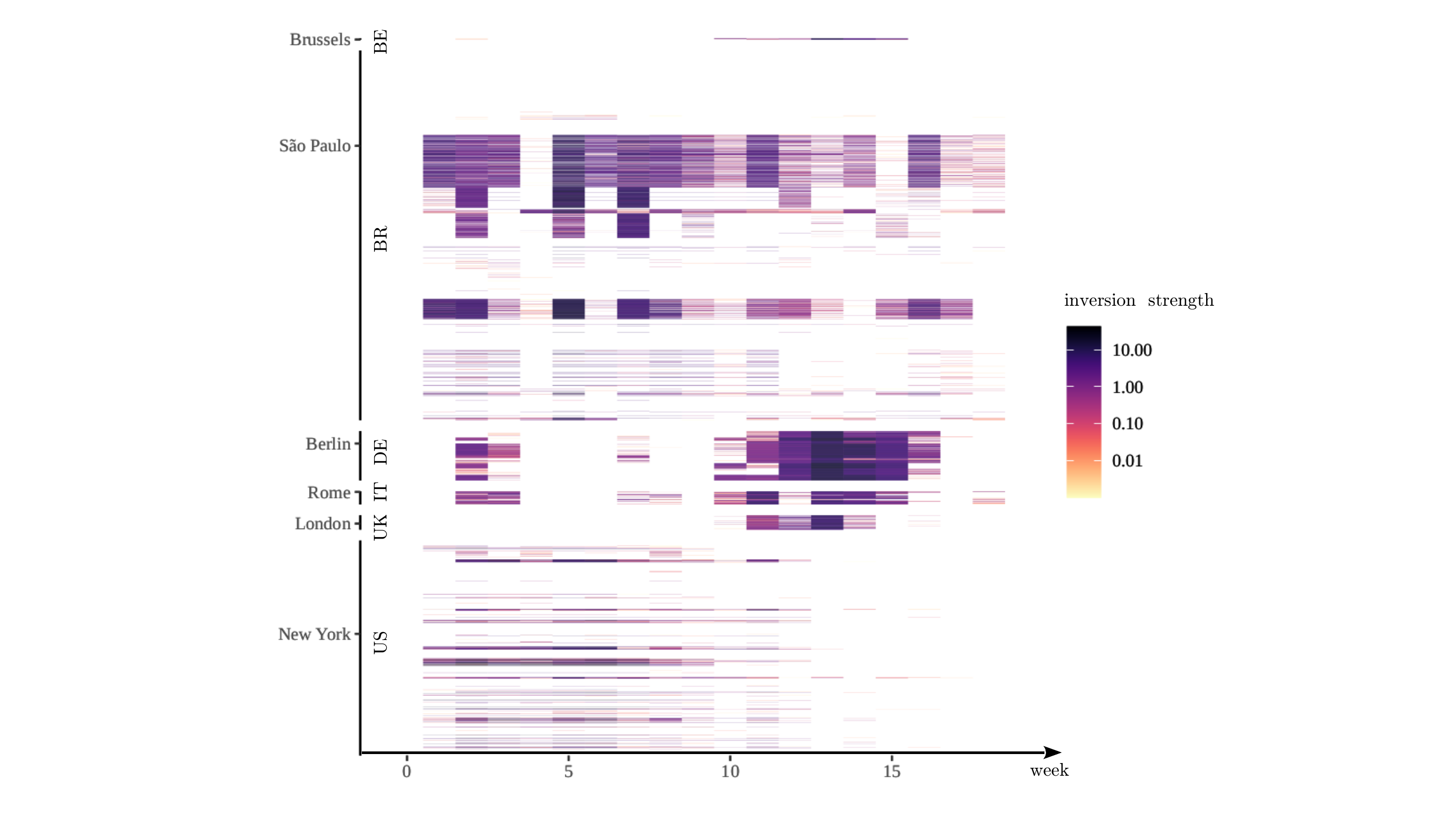}
                \end{center}
\scriptsize{
The figure illustrates the frequency of inversion episodes. The vertical axis refers to the districts. Countries and capital cities are labeled. The horizontal axis refers to the weeks. The color scale on the right indicates the strength of inversions. Weeks without inversion episodes are in white.}\\
        \label{fig:inverions_hm}
    \end{figure}
\FloatBarrier
\noindent We also compile data on total precipitation, downward UV radiation, specific humidity, air temperature 2 metres above the surface and U and V wind components from ERA5. Following \textcite{he2020}, we construct an indicator for still air from the wind component data, which is equal to one for wind speed no greater than 1 m/s.\\

\FloatBarrier
\subsection{Air pollution}
We use Aerosol Optical Depth (AOD) derived from MODIS satellite-images to proxy for $PM_{2.5}$ concentration in the atmosphere. We use the MCD19A2 V6 data product provided by NASA \autocite{lyapustin2018}. It measures AOD (blue band 0.47 $\mu$m) over land at a 1 kilometer resolution with global coverage combining images from the Terra and Aqua satellites. The data set can be used on Google Earth Engine. A link to the main script used to extract AOD values for our study units can be accessed here: \url{https://code.earthengine.google.com/ac966bffe655ba22379698d5709d9163}\\

We extract daily mean values of AOD pixels whose centroid lies in one of the districts, for which we have COVID-19 data. Note that Google Earth Engine's $.reduceRegions$ function does not extract values from pixels that do not have their centroid in the polygon of interest and therefore only works well with high resolution data, if study areas are small. For most longitudes both the Terra and the Aqua satellite pass over at around the same time during the day with a difference of around $\pm$1.5 hours. Grid cells at latitudes higher than 50 degrees tend to have several overpasses per day. The time attribute on the images is registered in UTC. Since we have study areas from different continents, we convert the time attribute into local solar time to avoid systematic differences in the time of the day at which we measure AOD. We only use images with the best quality according to the AOD\_QA band excluding for instance measurements from areas with cloud cover.\\

Figure 2 shows the link between daily inversion strength and $PM_{2.5}$ concentration. For our regression analysis we aggregate the daily observations to the week level and average over three weeks for regressions explaining COVID-19 cases and over four weeks for deaths. The aggregation accounts for potentially lagged effects of inversions on air quality.

   \begin{figure}
   \centering
      \caption{Air pollution and inversion strength }
   \includegraphics[clip = 7 0 7 5, width = 0.9\textwidth]{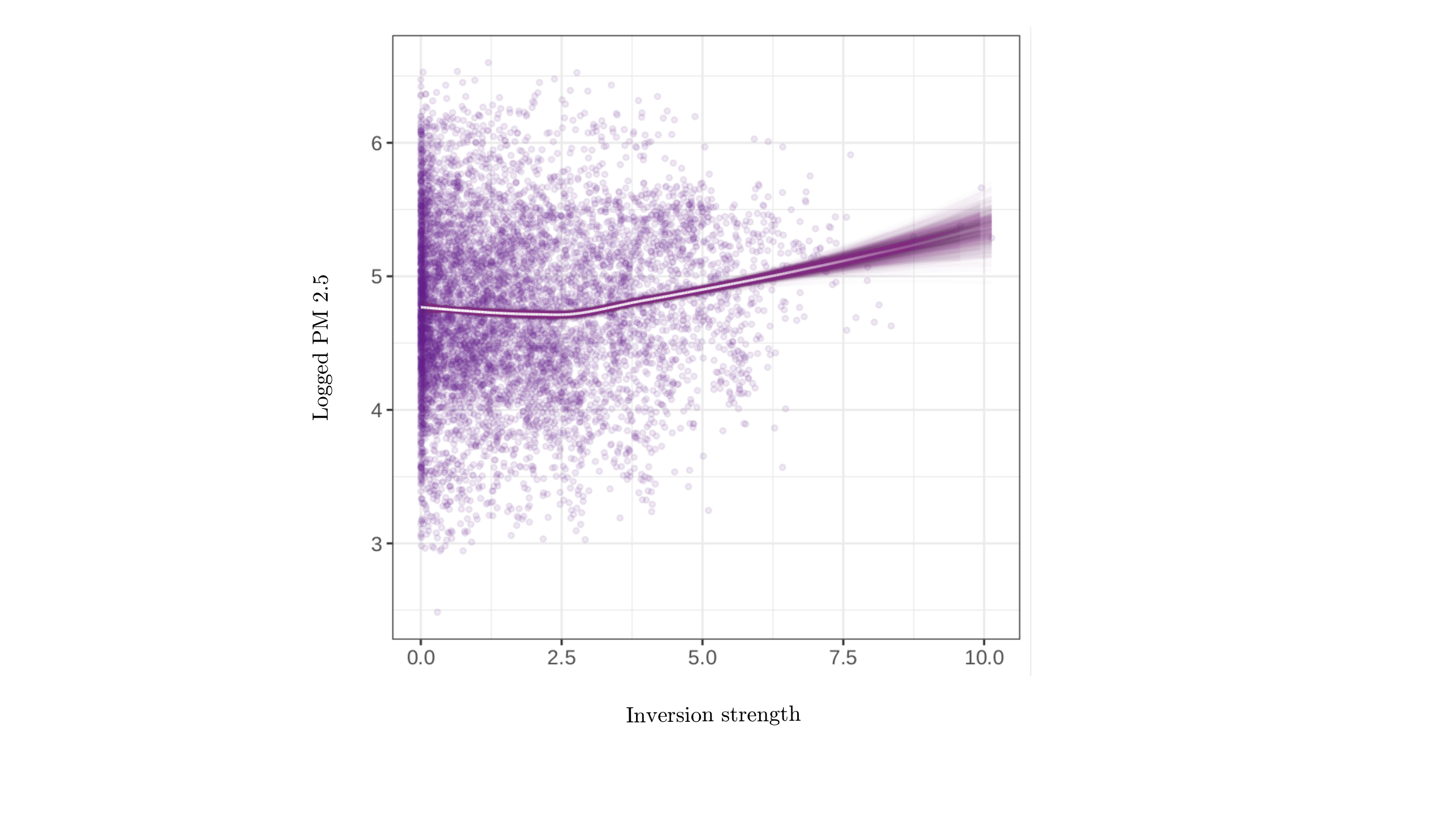}  
       \begin{flushleft}
       \vspace{-0.5cm}
    \scriptsize{
The figure shows a local polynomial regression fitting of the daily logged $PM_{2.5}$ observations determined by daily inversion strength. We bootstrap the fitting with 1000 repetitions. The white line represents the median of the fitted regression lines.}\\
    \end{flushleft}
\end{figure}

     \FloatBarrier

\subsection{Policy control variables}

\noindent In addition to the weather controls, we use data from the Oxford Covid-19 Government Response Tracker (OxCGRT)\autocite{oxcgrt}. The OxCGRT systematically collects information on countries' policy responses to the pandemic. It provides 17 indicators reflecting the extent of government action on a daily level. For our analysis we compose two control sets of these variables. The first one comprises containment and closure policies, namely, closings of schools and universities, closings of workplaces, cancellings of public events, closing of public transport, orders to "shelter-in-place" and otherwise confine to the home, and restrictions on internal movement between cities or regions. The second one comprises health system policies, namely, regulations on access to PCR testing and strategies for contact tracing after a positive diagnosis. To prevent high degrees of collinearity among the variables, we allow the sets to vary in size for the smaller sample in which we analyze COVID-19 deaths.\\

\subsection{Empirical approach}\label{ssec::gb}
    
\noindent The first-stage regression of the instrumental variable estimator is a generalized difference-in-differences regression model specified as:
\vspace{-0.2cm}
\begin{center}
\footnotesize{
\begin{align}
P_{iw} = \alpha_1 I_{iw} + \alpha_2 C_{iw-1} + W'_{iw} \alpha_3 +  M'_{iw} \alpha_4 +  T'_{iw} \alpha_5 + \gamma_{i} + \delta_{w} + \eta_{cm} + \epsilon_{iw}
\end{align}}
\end{center}

\noindent where the dependent variable is the logged average $PM_{2.5}$ concentration in district $i$ and week $w$ over the past three or four weeks. The parameter of interest, $\alpha_1$, represents the effect of inversions that accumulated over the same three or four weeks and that are weighted by their intensity, $I_{iw}$. We control for weather covariates $W'_{iw}$, COVID-related containment and closure measures $M'_{iw}$, and testing policies $T'_{iw}$. The variable $C_{iw-1}$ is the accumulated COVID-19 outcome, either cases or deaths, in the preceding week. The fixed effects $\gamma_{i}$, $\delta_{w}$, and $\eta_{cm}$ account for determinants of the dependent variable that are specific to each district $i$, each week $w$ and each month $m$ in state $s$.\\

\noindent In the second stage, the predicted $PM_{2.5}$ from equation 1 is used as an explanatory variable in
\vspace{-0.2cm}
\begin{center}
\footnotesize{
\begin{align}
C_{iw} = \beta_1 \widehat{P_{iw}} + \beta_2 C_{iw-1} + W'_{iw} \beta_3 +  M'_{iw} \beta_4 +  T'_{iw} \beta_5 + \gamma_{i} + \delta_{w} + \eta_{cm} + \mu_{iw}
\end{align}}
\end{center}

\noindent where $C_{iw}$ is the accumulated COVID-19 outcome of district $i$ in week $w$. The coefficient $\beta_1$ represents the percentage change in the dependent variable linked to a 1\%-increase in  inversion-driven air pollution $\widehat{P_{iw}}$. We estimate the second stage using Poisson pseudo-maximum likelihood estimation\autocite{correia2020}. We bootstrap standard errors clustered at the district level, where we assign treatment.\\

\noindent We also estimate a model at the daily level

\begin{center}
\footnotesize{
\begin{align}
C_{id} = \sum_{\theta}\rho_{\theta} I_{id}^{\theta} +
+ \rho_1 C_{id-1} + \lambda_{im} + \zeta_{sw} + \kappa_{cd} + \nu_{id}
\end{align}}
\end{center}

\noindent where the COVID-19 outcomes, $C_{id}$, and inversion variables, $I_{id}$, are now given for district $i$ and day $d$. We include $\theta$ lags of the inversion variable, where $\theta$ runs from -20 to 0 for regressions with cases as the dependent variable and from -27 to 0 for regressions with deaths as the dependent variable. Fixed effects $\lambda_{im}$, $\zeta_{sw}$, and $\kappa_{cd}$ absorb variation in $C_{id}$ that is specific to a district-month, state-week, and country-day.\\ 

Our instrumental variable approach can be conceptualized as a generalization of the difference-in-differences approach\autocite{goodman-bacon2018}. Our regressions form comparisons between groups that switch their treatment intensity on a given day and those that do not\autocite{dechaisemartind}. Similar to the standard difference-in-differences context a crucial assumption here is that both treatment and control group would have followed the same trend in the absence of treatment (common trends assumption). Note that treatment and control groups can be different at the baseline, but should be on the same trend. In our Poisson regressions we model the outcome as logarithmically distributed. Therefore, we assume that, after removing fixed effects and control variable variation, cases and deaths in districts hit by a shift in air pollution due to thermal inversion would have grown at the same rate as cases and deaths in districts that are not concerned by this natural experiment in a given week.\\
    
In the absence of random assignment into treatment, the main role of additional variables (controls and instruments) in our regression model is to eliminate reverse causality and omitted variable bias. For our estimate to capture a causal effect we have to rule out that a third confounding variable, such as mobility, economic activity or any other potentially non-measurable factor, is causally linked to both treatment (air pollution) and COVID-19 outcomes (infections and deaths) in either of three ways: (1) outcomes cause reductions in human activity causing reductions in air pollution (reverse causality), (2) human activity is driving both air pollution and outcomes (omitted variable bias), (3) air pollution and outcomes are both driving human activity (collider bias). In order to capture the net effect of air pollution we avoid introducing control variables that can also potentially be mediators or bad controls that are for instance influenced by our outcome and remove wanted variation if included in regressions. This approach is agnostic with regards to mechanisms and does not allow for an interpretation of variables beyond the treatment variable on which we focus our effort to restrict variation to quasi-experimental variation.\\
    
    % Our reduced-form approach keeps the model as parsimonious as possible and leverages exogenous factors that creates some random variation in the treatment conditional on a limited set of controls.
    
\subsection{Robustness checks}
\subsection*{Exclusion restriction}

\noindent The main threat to our identification strategy are violations of the exclusion restriction. This central assumption for our instrumental variable strategy would be violated if thermal inversions affected COVID-19 outcomes through other channels than air pollution. We conduct indirect tests to assess this hypothesis.\\

\noindent First, if people are aware of thermal inversions they might change their behavior, for instance, spend less time outside. This, in turn, could affect how the virus spreads. To mitigate this concern, we use nighttime-inversions. Nonetheless, we also analyze the effect of inversions on relative changes in movement and the time spent at different types of locations. We use movement data from the social media platform Facebook, which compiles information on people’s precise locations from their mobile devices\autocite{fb}. Data are available starting on 15th February 2020 at the district-level for Brazil and the United States and at the state-level for the European countries in our sample. For every country we use data at the highest resolution available and cluster standard errors respectively. The results in Table 5 do not point to causal effects of inversions on people’s activities. \\

  \begin{table}[!ht]
    \begin{center}
    \scriptsize{    \begin{tabularx}{0.85\textwidth}{@{}p{4cm}
    S[table-format=2.4]S[table-format=2.4]S[table-format=2.4]@{}}
   \hline
& {Relative change} & & {Proportion staying within} \\
& {in movement} & & {a single location}\\
&{(1)} & & {(2)} \\
\hline
inversion strength &0.0010  && -0.0000\\
  &{(}0.0007{)} &&{(}0.0003{)}\\ [0.2cm]
\hline
Countries& \multicolumn{3}{c}{\footnotesize{BEL, BRA,DEU, GBR, ITA, USA}}\\
\hline
\end{tabularx}}
    \end{center}
\setlength{\leftskip}{1.2cm}
\setlength{\rightskip}{1.2cm}
\scriptsize{
The regression includes district-month, state-week and country-day fixed effects. We control for weather covariates. Standard-errors clustered at the district/state level are in parentheses. The sample size is 276,418. Signif. Codes: ***: 0.01, **: 0.05, *: 0.1}

\setlength{\leftskip}{0pt}
\setlength{\rightskip}{0pt}

\caption{The effect of thermal inversions on movement}
\end{table}

  \FloatBarrier
  
\noindent Second, we provide reduced form estimates that represent the overall effect of thermal inversions on COVID-19 outcomes. If inversions have no effect on COVID-19 outcomes other than through air pollution, we would expect that the overall effect of inversions on COVID-19 outcomes is approximately equal to the product of the effect of inversion-driven air pollution on COVID-19 outcomes (Table 2) and the effect of inversions on air pollution (Table 1). The results in Table 6 confirm this hypothesis. For instance, when multiplying the coefficient in column (3) in Table 1 with the corresponding coefficient in Table 2, we obtain a value of 0.0086 which is close to the estimate in column (3) in Table 6.\\

  \begin{table}[!ht]
    \centering
\scriptsize{
    \begin{tabularx}{\textwidth}{@{}p{4cm}
    S[table-format=2.4]S[table-format=2.4]S[table-format=2.4]|S[table-format=2.4]S[table-format=2.4]S[table-format=2.4]@{}}
   \hline
& \multicolumn{3}{c}{\textbf{COVID-19 cases}}& \multicolumn{3}{c}{\textbf{COVID-19 deaths}}\\
& \multicolumn{3}{c}{three-week window}& \multicolumn{3}{c}{four-week window}\\
&{(1)} & {(2)} & {(3)} & {(4)} & {(5)} & {(6)}\\
\hline
\textbf{inversion strength} &0.0000  &0.0079{$^{***}$}&0.0088{$^{***}$}&0.0105{$^{**}$}&0.0142{$^{***}$}&0.0167{$^{***}$}\\
& {(}0.0017{)}  &{(}0.0013{)}  &{(}0.0014{)} &{(}0.0050{)} &{(}0.0034{)}  &{(}0.0028{)} \\[0.2cm]
\hline
Observations &{72,021} & {72,021} & {72,021}& {20,658}& {20,658}& {20,658}\\
Countries& \multicolumn{3}{c}{\footnotesize{BEL, BRA, DEU, GBR, ITA, USA}}& \multicolumn{3}{c}{\footnotesize{BEL, BRA, DEU, USA}}\\
\hline
weather controls & {yes}& {yes}& {yes}& {yes}& {yes}& {yes}\\
containment controls  & & {yes}& {yes}& & {yes}& {yes}\\
health system controls & & & {yes}& & & {yes}\\
\hline
\end{tabularx}}
\justify{
\scriptsize{
In each panel control variables are added sequentially from left to right. The first set of controls contains weather variables only, the second set adds controls for COVID19-related containment and closure policies (e.g. school closings and stay at home requirements), and the third set additionally adds COVID19-related health system policies (e.g. testing policies and contact tracing). In the left panel we control for the accumulated case numbers of the preceding week, in the right panel we control for accumulated death number of the preceding week. All regressions include district, week and state-month fixed effects. Standard-errors clustered at the district level are in parentheses. Signif. Codes: ***: 0.01, **: 0.05, *: 0.1}}
\caption{The effect of thermal inversions on COVID-19 outcomes}
\end{table}
\FloatBarrier

\noindent Third, we expand our control set of weather covariates. Inversion episodes are correlated with weather conditions. At the same time, local weather conditions strongly correlate with how people spend their time and, thus, may affect infections. In our main analysis we control for precipitation, humidity, temperature and still air. 
In Figure 3, we present estimates based on 14 alternative sets of controls that allow for more complex relations among the weather variables. The upper panels present the effect of inversions on logged $PM_{2.5}$ levels, while the lower panels present their effect on COVID-19 cases and deaths. The estimates highlighted in purple are our baseline results from Table 1 and Table 6. Given that the coefficients in all four panels are largely insensitive to the choice of weather controls, we believe that the exclusion restriction is likely to hold.\\

\begin{figure}[!ht]
    \centering 
    \caption{Alternative weather controls}
        \centering 
       \includegraphics[trim = 0 0 6cm 0 , width = 1.1\textwidth]{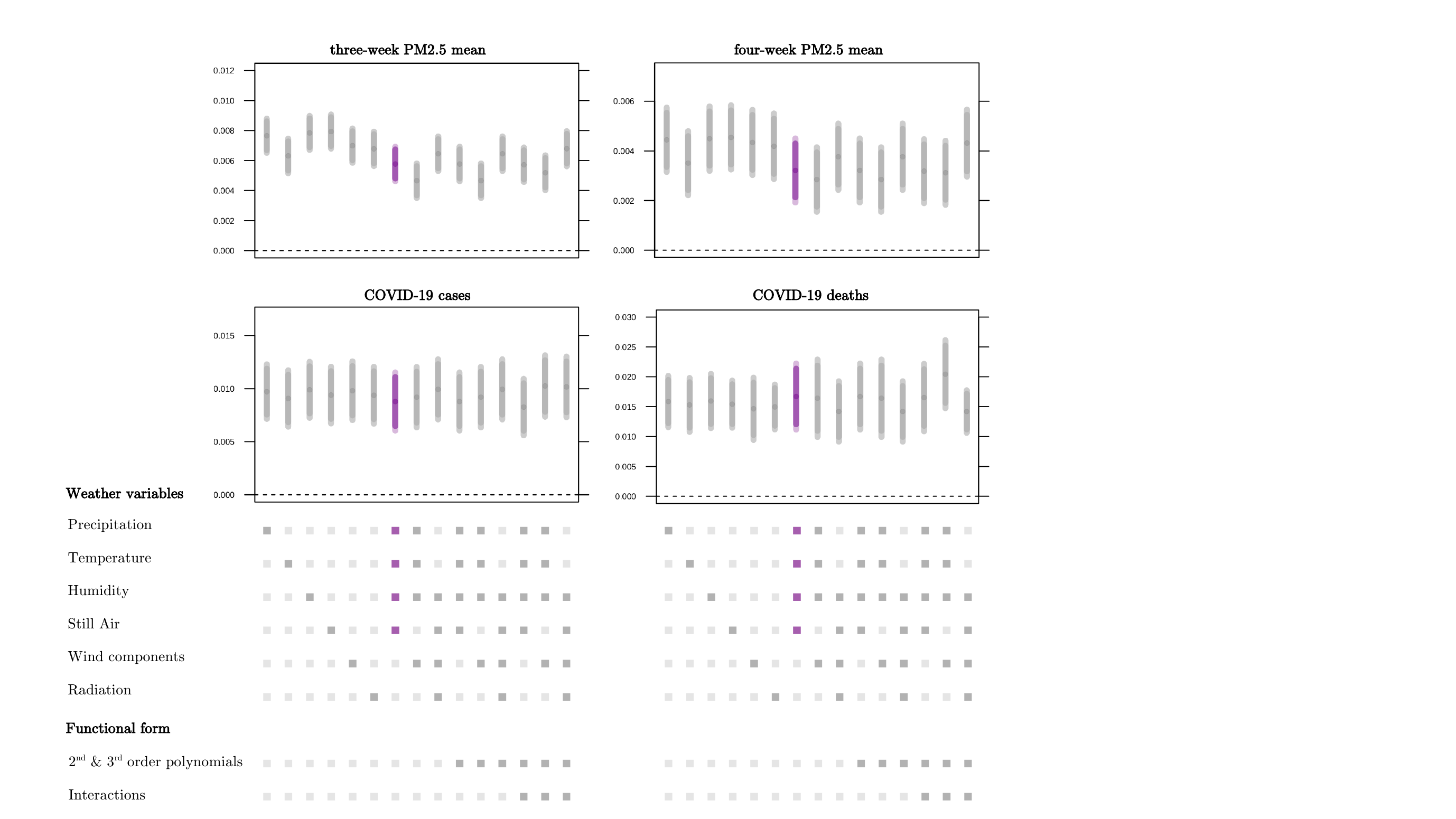}
    \label{fig:weather}
    \begin{flushleft}
    \scriptsize{
The four panels in the figure represent the effect of thermal inversions on the dependent variable stated above each panel. In each panel the coefficients are estimated using different weather control sets which are specified by the specification chart below. All regressions include containment and health system controls. Confidence intervals are plotted for the 5 and 10\% level of statistical significance.}\\
    \end{flushleft}
\end{figure}

 \FloatBarrier
  
 \subsection*{Spillovers}  
 \noindent Another assumption of our identification strategy is that air pollution effects on COVID-19 outcomes do not spill across district borders. This is called the stable unit treatment value assumption (SUTVA). However, this assumption may be violated given the nature of a pandemic outbreak. Infected people usually travel freely between districts and may thereby increase the infections in neighboring districts as well. With regards to our analysis this means, that inversions may not only affect COVID-19 outcomes in the same district but also in the surrounding ones. In this case, the estimated effects in Table 6 would be underestimated. To assess the extent of a potential bias linked to cross-district spillover effects, we restrict our sample to districts and weeks that are subject to a restrictive lock-down. We consider only districts in which people are required to stay at home with exceptions for daily exercise, grocery shopping, and essential trips. As these containment measures are specifically targeted to prevent virus transmission beyond a small radius around an individual's home, we assume that spillover effects are less of a concern in this sample. Table 8 reproduces the regression coefficients of Table 6. We only control for weather covariates, because the districts remaining in this sub-sample are very similar in terms of containment strategies and  control variables for political measures are therefore highly collinear. Compared to the inversion effect on COVID-19 case numbers in Table 6 (0.0088) the estimated coefficient in Table 8 is slightly larger in magnitude. For COVID-19 deaths, however, coefficients are similar in magnitude. We conclude that if spillover effects exist, they are likely to make our main results lower-bound estimates.\\
 
  \begin{table}[!ht]
    \centering
\begin{center}
    \scriptsize{    \begin{tabularx}{0.85\textwidth}{@{}p{4cm}
    S[table-format=2.4]S[table-format=2.4]S[table-format=2.4]@{}}
    \hline
& {\textbf{COVID-19 cases}}& {\textbf{COVID-19 deaths}}\\
& {three-week window}& {four-week window}\\
&{(1)} & {(2)} \\
\hline
inversion strength & 0.0134{$^{***}$}&0.0158{$^{***}$}\\
  &{(}0.0013{)} &{(}0.0030{)}\\ [0.2cm]
\hline
Observations & {13,885}&  {5,145}\\
Countries& \multicolumn{1}{c}{\footnotesize{BEL, DEU, GBR, ITA, USA}}& \multicolumn{1}{c}{\footnotesize{BEL,  DEU, USA}}\\
\hline
\end{tabularx}}
\end{center}

\setlength{\leftskip}{1.1cm}
\setlength{\rightskip}{1.1cm}
\scriptsize{
In the left panel we control for the accumulated case numbers of the preceding week, in the right panel we control for accumulated death number of the preceding week. All regressions include district, week and state-month fixed effects as well as weather controls. Standard-errors clustered at the district level are in parentheses. Signif. Codes: ***: 0.01, **: 0.05, *: 0.1\\}

\setlength{\leftskip}{0pt}
\setlength{\rightskip}{0pt}

\caption{The effect of thermal inversions on COVID-19 outcomes - lock-down sample.}
\end{table}
\FloatBarrier
  
%     - Inversion effect on google mobility
%     - Other weather controls
%     - Policy and testing variables
%     - Reduced Form with Facebook Data

%     \paragraph{Heterogeneity}
%     - Different country samples
%     - Other FEs
%     \paragraph{OLS}
%     \paragraph{SUTVA, spillovers from}
%     - Subsample with mandatory stay-at-home requirements
%     \paragraph{Different time lags}
%     - 28 lags for cases, 35 for deaths
\newpage
\subsection{Data and code availability}

%The datasets generated during and/or analysed during the current study are available at https://github.com/hannahklauber/cov19_pollution with the identifier  10.5281/zenodo.3972186

Data and code are available in the following github repository:\\

\url{https://github.com/hannahklauber/cov19_pollution}\\

DOI: 10.5281/zenodo.3973516 \\

\printbibliography
\end{refsection}

%\textbf{Acknowledgements:} We thank Jeffrey Wooldridge for advice on our two-stage procedure. We thank Yujie Wang for advice with the air pollution data.\\

%\textbf{Author contributions:} 
%\textbf{Competing interests:} The authors declare no competing interests.\\
%\textbf{Correspondence and requests for materials} should be addressed to 
\end{document}